\documentclass[10pt,twocolumn,aps,prd,nofootinbib]{revtex4-2}
\usepackage{amsmath,amssymb,amsfonts,mathtools}
\usepackage{bm}
\usepackage{hyperref}
\usepackage{physics}
\usepackage{graphicx}
\usepackage{enumitem}
\newcommand{\Lie}{\mathcal{L}}
\newcommand{\cO}{\mathcal{O}}

\newcommand{\ept}{\bm{\epsilon}_2}

\newcommand{\epf}{\bm{\epsilon}_4}

\begin{document}

\title{Extended black hole thermodynamics in a DGP braneworld}
\author{Naman Kumar}
\email{namankumar5954@gmail.com}
\affiliation{Department of Physics,
Indian Institute of Technology Gandhinagar, Palaj, Gujarat, India, 382355}

\begin{abstract}
We develop extended black-hole thermodynamics on a Dvali--Gabadadze--Porrati (DGP) brane by promoting the brane tension \(\sigma\) to a thermodynamic variable within the extended Iyer--Wald framework. The brane tension acts as a localized vacuum energy with pressure \(P_\sigma \equiv -\sigma\), yielding a new work term \(V_\sigma\,\mathrm{d}P_\sigma\) in the first law and the corresponding Smarr relation. For static, spherically symmetric black holes we show that the conjugate volume equals the geometric volume \(V_\sigma=\tfrac{4\pi}{3}r_h^3\); for stationary, axisymmetric solutions it admits a covariant, slice-independent definition and evaluates to \(V_\sigma=\tfrac{4\pi}{3}\!\left(r_+^3+a^2 r_+\right)\). Working on the ghost-free normal branch, the brane is asymptotically flat with a single horizon, so the construction avoids de Sitter obstructions. Along a flat-brane path, asymptotic flatness is preserved by co-varying the bulk cosmological constant, and induced-gravity effects are suppressed by \(r_h/r_c\). These results establish a consistent flat-braneworld realization of black-hole chemistry in which brane tension provides the physically motivated pressure variable.
\end{abstract}

\maketitle

\section{Introduction}

For AdS black holes, pressure is identified with the cosmological constant $\Lambda$, giving rise to the extended black hole thermodynamics \cite{Dolan:2010ha,Kastor:2009wy,Kubiznak:2012wp} or black hole chemistry \cite{Wei:2015iwa,Kubiznak:2016qmn,Wei:2019uqg,Frassino:2015oca}. The identification is physically meaningful since the pressure is given as
\begin{equation}
    P=-\frac{\Lambda}{8\pi},
\end{equation}
which is naturally positive as $\Lambda$ is negative for an AdS black hole. This suggests that extended black hole thermodynamics is unique to AdS black holes. Interestingly, this identification of thermodynamic pressure and volume gives rise to the van der Waals type phase transition for black holes \cite{Chamblin:1999tk,Kubiznak:2012wp}.\vspace{2mm}

In de Sitter (dS) spacetime with $\Lambda>0$, defining a thermodynamic pressure analogous to the AdS case is inherently problematic. The primary obstruction arises from the \emph{absence of a global timelike Killing vector}: while asymptotically AdS spacetimes possess a timelike Killing field extending to spatial infinity, enabling a Hamiltonian formulation with well-defined conserved charges, dS spacetime does not. Its Killing vectors become spacelike at large distances, and the spacetime terminates at future and past \emph{spacelike} boundaries rather than at a timelike infinity. Consequently, standard notions of energy and mass---ADM, Komar, or holographic---are not globally well-defined.

A second obstruction is the \emph{horizon structure} of dS spacetimes. Generic dS black holes possess both a black hole horizon and a cosmological horizon, each with distinct surface gravities and hence different temperatures. This prevents the definition of a single, global thermal equilibrium state. While one can formally associate a ``pressure'' $P=-\Lambda/8\pi<0$ with the positive cosmological constant, the lack of a consistent global energy and the coexistence of multiple, inequivalent temperatures mean that the usual extended first law,
\begin{equation}
dM = T\,dS + V\,dP + \dots,
\end{equation}
cannot be implemented in a physically meaningful or observer-independent manner. Any thermodynamic description becomes patch-dependent, undermining its universality.

Our task is to show that braneworlds allow for a more natural setting for extended black hole thermodynamics, where the above problems, being confined to asymptotic AdS and difficulty in extending to dS spacetimes, are alleviated by trading the brane tension $\sigma$ with the cosmological constant $\Lambda$. We show this using the Iyer--Wald formalism \cite{Wald:1993nt,Iyer:1994ys} whose extended version has been recently proposed to derive the extended black hole thermodynamics \cite{Xiao:2023lap}. Very recently, Iyer--Wald formalism has been applied to the restricted phase space thermodynamics to obtain the Euler relation for black hole systems \cite{Chen:2025tdt}. Moreover, Frassino et al. \cite{Frassino:2022zaz} 
used braneworld holography to argue that varying the brane tension induces 
a dynamical cosmological constant on the brane, thereby providing a higher--dimensional origin of extended black hole thermodynamics.\footnote{In the pure tension braneworld considered in \cite{Frassino:2022zaz}, varying $\tau$ is equivalent to varying the on--brane cosmological constant, see their eq.\,(9). 
However, this identification is model--dependent. As they themselves note 
(see their eqs.\,(31)--(32)), once a DGP term is included on the brane, the 
relation between $\tau$ and $\Lambda_d$ is decoupled, allowing one to work 
in an ``isocosmological DGP ensemble'' where $\Lambda_d$ is held fixed while 
$\tau$ is varied. It is precisely this setting in which we carry out our analysis.} 
While their approach motivates tension--variation holographically, our analysis 
takes a different route: we consider a DGP braneworld and use the Iyer--Wald 
formalism to derive the extended first law directly, keeping the cosmological 
constant fixed (including the flat--brane case). In this way, we establish 
that extended black hole thermodynamics arises naturally in a DGP braneworld, 
independent of AdS/CFT considerations.

\subsection*{Positive Brane Tension in Braneworld Setups}

In braneworld scenarios such as Randall--Sundrum \cite{Randall:1999vf,Randall:1999ee} and DGP models \cite{Dvali:2000hr}, the sign of the brane tension $\sigma$ is crucial. A \emph{positive-tension brane} supports a normalizable graviton zero mode and yields a well-defined four-dimensional Newton constant on the brane. It satisfies the null energy condition and avoids ghost-like instabilities that plague \emph{negative-tension branes}, where fluctuations typically carry negative kinetic energy. In the DGP model specifically, the positive-tension branch corresponds to the so-called \emph{normal branch}, which is stable and ghost-free, while the self-accelerating branch (effectively associated with negative tension) exhibits ghost instabilities. Thus, positive brane tension is the only physically viable choice for a stable braneworld.

\subsection*{Negative Pressure from Positive Brane Tension in DGP}

The stability and asymptotic structure of a positive-tension brane enable a consistent definition of pressure and extended thermodynamics in the DGP setup. A positive-tension brane induces a vacuum energy density $\rho_\sigma=\sigma$ on the brane, whose stress-energy tensor reads
\begin{equation}
T_{\mu\nu}^{(\sigma)} = -\sigma\,h_{\mu\nu},
\end{equation}
implying a \emph{negative pressure} $p_\sigma=-\sigma$. Crucially, this vacuum energy is \emph{localized} to the brane rather than filling the entire bulk spacetime. The brane is asymptotically flat in the normal branch, possessing a well-defined spatial infinity and a global timelike Killing vector in the exterior of a localized black hole. These properties allow one to construct ADM charges and implement the Iyer--Wald formalism in a straightforward manner. As a result, one obtains a well-defined extended first law,
\begin{equation}
dM = T\,dS + V_\sigma\,dP_\sigma + \dots, \qquad P_\sigma=-\sigma,
\end{equation}
with a corresponding Smarr relation,
\begin{equation}
M = 2TS + 2\Omega J + \Phi Q - 2V_\sigma P_\sigma.
\end{equation}
Here, the negative pressure $P_\sigma$ enters on exactly the same footing as the positive pressure $P=-\Lambda/8\pi$ in AdS, but without the conceptual difficulties of de Sitter: there is a single horizon, a single temperature, and a well-defined notion of energy.

In summary, while de Sitter spacetime fails to admit a globally consistent thermodynamic pressure due to its causal structure and multiple horizons, a DGP braneworld with \emph{positive brane tension} yields a stable, ghost-free setting in which the induced negative pressure is well-defined and directly participates in a consistent extended black hole thermodynamics.

\subsection*{Motivation for our approach}

There are two key motivations behind our study:\vspace{1mm}

\paragraph*{\bf (i) Brane tension versus cosmological constant.}
In conventional extended black hole thermodynamics, the
cosmological constant $\Lambda$ is promoted to a thermodynamic
variable, with $P=-\Lambda/8\pi$ identified as the pressure.
While mathematically consistent, this identification is conceptually
less direct: $\Lambda$ is usually a fixed parameter of the action
rather than a genuine physical condensate that can fluctuate.  
By contrast, in braneworld models the \emph{brane tension} $\sigma$
is a well-defined localized energy density confined to the brane.
It directly sources the brane Einstein equations as $T^{(\sigma)}_{\mu\nu}
=-\sigma g_{\mu\nu}$, providing an intrinsic notion of vacuum energy
for observers on the brane. Moreover, positive brane tension is
dynamically motivated: it supports a normalizable graviton zero mode,
satisfies the null energy condition, and avoids the ghost
instabilities associated with negative-tension or self-accelerating
branches. Thus, varying $\sigma$ is physically more meaningful than
varying a bulk $\Lambda$: it corresponds to varying a brane-localized
vacuum energy, and it naturally yields a negative pressure
$P_\sigma=-\sigma$ that participates in the first law.  Moreover, Braneworld setups with varying brane tension $\sigma$ have already been studied in the context of cosmology in the literature \cite{Gergely:2008fw,Wong:2010rg,Garcia-Aspeitia:2018fvw,Kumar:2024Variable}, which further strengthens our motivation to treat it as a varying parameter.

\paragraph*{\bf (ii) Beyond asymptotically AdS.}
A second motivation is that our framework extends the scope of extended
thermodynamics beyond the AdS context. The AdS case is privileged
because $\Lambda<0$ yields a single global timelike Killing vector at
infinity, ensuring a well-defined ADM mass and a unique Hawking
temperature. In de Sitter, however, multiple horizons with different
temperatures obstruct a consistent first law, and a global conserved
energy is lacking. By contrast, in the DGP braneworld with positive
tension, the normal branch is asymptotically flat: it possesses a
timelike Killing field at infinity, a single black hole horizon with
temperature $T$, and well-defined ADM charges. This allows the Iyer--Wald
construction to go through cleanly. Hence, our proposal shows that
extended black hole thermodynamics, often regarded as an ``AdS-only'' phenomenon, arises naturally on a flat brane background once brane
tension is treated as a thermodynamic variable.

\medskip
In this way, our work provides a new setting for black hole chemistry:
by trading the bulk cosmological constant for a brane-localized vacuum
energy, we obtain a consistent and physically motivated extended first
law even in asymptotically flat spacetimes.\vspace{2mm}

The remainder of the paper is organized as follows. In Sec.~\ref{sec_II}, we
introduce the DGP action with brane tension and discuss the junction
conditions governing the brane dynamics. In Sec.~\ref{sec_III}, we formulate the
extended Iyer--Wald construction with variable brane tension, identifying
the new work term associated with the $\sigma$--variation. In Sec.~\ref{sec_IV}, we
derive the extended first law and Smarr relation on the brane. In Sec.~\ref{sec_V},
we present an explicit computation for a static, spherically symmetric
black hole on the brane, showing how the $\sigma$--volume reduces to the
geometric volume $V_\sigma=4\pi r_h^3/3$. In Sec.~\ref{sec:rotating-case}, we generalize our analysis to stationary, axisymmetric black holes on the brane 
and show that the conjugate thermodynamic volume $V_\sigma$ admits a covariant, slice-independent definition even in the rotating case. For a Kerr seed geometry, we obtain an explicit closed-form expression $V_\sigma = \frac{4\pi}{3}(r_+^3 + a^2 r_+)$, recovering the static result in the non-rotating limit and demonstrating that our framework remains robust when angular momentum is included. In Sec.~\ref{sec_VI}, we address how to maintain asymptotic flatness along the family of solutions by co-varying
the bulk cosmological constant. We conclude with a summary of results and
potential extensions in Sec.~\ref{sec_VII}. \vspace{2mm}

In Appendix~\ref{app_A}, we show that the $\cO(r_h/r_c)$ contribution to both the thermodynamic volume and the canonical mass vanishes on the flat-brane path, so the extended first law holds unmodified up to $\cO((r_h/r_c)^2)$.

\section{DGP action with brane tension and boundary terms}
\label{sec_II}
Consider a $5$D bulk $\mathcal{M}_5$ with a timelike codimension-1 brane $\mathcal{B}$ (our universe). The DGP action with a brane tension $\sigma$ is
\begin{align}
    S &= \frac{M_5^3}{2}\int_{\mathcal{M}_5}\dd^5x\,\sqrt{-g_5}\,(R_5-2\Lambda_5)
    + 2M_5^3\int_{\mathcal{B}}\dd^4x\,\sqrt{-g_4}\,K \nonumber\\
    &\quad + \int_{\mathcal{B}}\dd^4x\,\sqrt{-g_4}\!\left(\frac{M_4^2}{2}R_4 - \sigma + \mathcal{L}_{\rm matter}\right),
    \label{eq:DGPaction}
\end{align}
where $K$ is the trace of the extrinsic curvature of $\mathcal{B}$ (the factor of $2$ assumes a $\mathbb{Z}_2$-symmetric embedding; for a one-sided brane one uses the standard Gibbons--Hawking coefficient), and $R_4$ is built from the induced metric $g_{\mu\nu}$ on $\mathcal{B}$.

Varying \eqref{eq:DGPaction} yields the brane field equation in ``junction'' form,
\begin{equation}
    M_4^2\,G_{\mu\nu}
    = T_{\mu\nu} - \sigma\,g_{\mu\nu}
    + 2M_5^{3}\!\left(K_{\mu\nu}-K g_{\mu\nu}\right),
    \label{eq:junction}
\end{equation}
with $T_{\mu\nu}$ the matter stress-energy localized on $\mathcal{B}$.\vspace{2mm}

\paragraph*{\bf Vacuum background and asymptotics:}
For a spatially flat FRW brane, the vacuum ($\rho=0$) Friedmann equation is
\begin{equation}
    H^2 - \frac{\epsilon}{r_c} H
    = \frac{\sigma}{3M_4^2} + \frac{\Lambda_5}{6} + \frac{C}{a^4},\qquad \epsilon=\pm1,
    \label{eq:FriedmannDGP}
\end{equation}
where, $r_c=M_4^2/M_5^3$ is the crossover scale between 4D and 5D gravity, $\epsilon=\pm1$ corresponds to the normal and self-accelerating branch respectively, and $C/a^4$ is the dark radiation constant. A \emph{flat} brane ($H_0=0$) requires the tuning
\begin{equation}
    \frac{\sigma}{3M_4^2}+\frac{\Lambda_5}{6}=0
    \;\;\Longleftrightarrow\;\;
    \sigma = -\frac{1}{2}M_4^2\,\Lambda_5.
    \label{eq:flatTuning}
\end{equation}
Otherwise, the asymptotics are (A)dS with $H_0\neq 0$, determined by \eqref{eq:FriedmannDGP}.

\section{Iyer--Wald formalism with a variable brane tension}
\label{sec_III}
Let $L(\phi;\lambda)$ be a diffeomorphism-covariant Lagrangian $4$-form on the brane depending on fields $\phi$ and a constant parameter $\lambda$. The variation reads
\begin{equation}
    \delta L = E(\phi)\,\delta\phi + \dd\Theta(\phi;\delta\phi) + \frac{\partial L}{\partial \lambda}\,\delta\lambda,
    \label{eq:genvar}
\end{equation}
with Euler--Lagrange form $E=0$ on-shell and symplectic potential $3$-form $\Theta$. For a vector field $\xi$, we define the Noether current as
\begin{equation}
    J_\xi \equiv \Theta(\phi;\Lie_\xi\phi) - \xi\cdot L.
\end{equation}
On-shell, $J_\xi$ is closed and can be written as $J_\xi=\dd Q_\xi$, where $Q_\xi$ is the Noether charge $2$-form. The Hamiltonian variation associated with the symmetry generated by $\xi$ is
\begin{equation}
    \delta H_\xi
    = \int_{\partial\Sigma}\!\left(\delta Q_\xi - \xi\cdot\Theta\right)
      - \int_{\Sigma}\xi\cdot\left(\frac{\partial L}{\partial\lambda}\right)\delta\lambda,
    \label{eq:MasterVari}
\end{equation}
for a Cauchy slice $\Sigma$ with boundary $\partial\Sigma$ (typically the union of a horizon cross-section and an outer boundary).\vspace{2mm}

We now apply \eqref{eq:MasterVari} to the \emph{brane} Lagrangian $4$-form
\begin{equation}
    L_{\mathcal{B}} = \frac{M_4^2}{2}\,R_4\,\epf - \sigma\,\epf + L_{\rm matter}[\phi]\,,
    \label{eq:Lbrane}
\end{equation}
treating $\sigma$ as the variable coupling $\lambda$. (The bulk terms enter through the junction condition \eqref{eq:junction}; in the regime $r_h\ll r_c$ they induce $\cO(r_h/r_c)$ corrections that we track explicitly below.)

From \eqref{eq:Lbrane} we have
\begin{equation}
    \frac{\partial L_{\mathcal{B}}}{\partial \sigma} = -\,\epf.
\end{equation}
Thus, for a stationary configuration with Killing field $\xi^\mu$ normalized to unit at infinity, the ``work'' term from varying $\sigma$ is
\begin{equation}
    -\int_{\Sigma}\xi\cdot\left(\frac{\partial L_{\mathcal{B}}}{\partial \sigma}\right)\tilde\delta\sigma
    = \int_{\Sigma}\xi\cdot\epf\,\tilde\delta\sigma
    \equiv V_\sigma\,\tilde\delta\sigma,
    \label{eq:VsigmaDef}
\end{equation}
where $\tilde\delta$ allows for the variation of brane tension $\sigma$, in which case we define the metric variation as
\begin{equation}
   \tilde\delta g_{\mu\nu}=\delta g_{\mu\nu}+\frac{\partial g_{\mu\nu}}{\partial\sigma}\tilde\delta\sigma
\end{equation}
and the \emph{$\sigma$-volume} $V_\sigma$ is the (background-subtracted) proper $3$-volume on the brane Cauchy slice,
\begin{equation}
    V_\sigma \;\equiv\; \lim_{R\to\infty}\left[
    \int_{\Sigma\cap\{r\le R\}}\!\!\!\!\dd^3x\,\sqrt{h}
    \;-\;\int_{\Sigma_{\rm ref}\cap\{r\le R\}}\!\!\!\!\dd^3x\,\sqrt{h_{\rm ref}}
    \right].
    \label{eq:VsigmaSubtraction}
\end{equation}
Here $h_{ij}$ is the induced $3$-metric on $\Sigma$, and the reference background depends on the path in parameter space:
\begin{itemize}[leftmargin=1.25em]
    \item[(A)] \textbf{Unconstrained $\sigma$-variation:} we take $\Sigma_{\rm ref}$ to be the vacuum with the \emph{same} $\sigma$ (AdS/dS/Minkowski as dictated by \eqref{eq:FriedmannDGP}).
    \item[(B)] \textbf{Flat-brane path:} we enforce $H_0=0$ along the variation by co-varying $\Lambda_5$ (see Sec.~\ref{sec_VI}). Then the reference is Minkowski on the brane.
\end{itemize}

For the Einstein--Hilbert part $(M_4^2/2)R_4\epf$, the Noether charge and symplectic potential are~\cite{Iyer:1994ys}
\begin{align}
    (Q_\xi)_{\mu\nu} &= -\frac{M_4^2}{2}\,\nabla_{[\mu}\xi_{\nu]}\,\ept^{\mu\nu},\\
    \Theta^\mu(\tilde\delta g) &= \frac{M_4^2}{2}\left(\nabla_\nu \tilde\delta g^{\mu\nu} - \nabla^\mu \tilde\delta g\right),
\end{align}
with $\ept$ the binormal area 2-form on the relevant codimension-2 surface and $\tilde\delta g\equiv g_{\alpha\beta}\tilde\delta g^{\alpha\beta}$. As usual, the surface integral of $\delta Q_\xi - \xi\cdot\Theta$ over the bifurcation surface gives $T\tilde\delta S$, while the integral at infinity gives $\tilde\delta M - \Omega\,\tilde\delta J - \Phi\,\tilde\delta Q$ (if rotation and gauge fields are present), all with Newton constant $G_4$.

\section{First law with variable brane tension}
\label{sec_IV}
Putting \eqref{eq:MasterVari} and \eqref{eq:VsigmaDef} together for a stationary black hole spacetime on the brane,
\begin{equation}
    \tilde\delta M
    = \frac{\kappa}{8\pi G_4}\,\tilde\delta A \;+\; \Omega\,\tilde\delta J \;+\; \Phi\,\tilde\delta Q
      \;+\; V_\sigma\,\tilde\delta\sigma \;+\; \cO\!\left(\frac{r_h}{r_c}\right).
    \label{eq:firstlaw}
\end{equation}
The $\cO(r_h/r_c)$ terms encode the leading induced-gravity corrections due to the extrinsic-curvature contribution in \eqref{eq:junction}. In the \emph{decoupling limit} $r_c\to\infty$, they vanish and \eqref{eq:firstlaw} reduces exactly to the GR first law with a variable vacuum term $-\sigma$ on the brane.\vspace{2mm}

It is convenient to define the pressure-like quantity
\begin{equation}
    P_\sigma \equiv -\,\sigma,
\end{equation}
with $\sigma>0$ as argued in the introduction, so that the new work term reads $-\,V_\sigma\,\tilde\delta P_\sigma$, in direct analogy with $V\,\tilde\delta P$ in AdS black hole chemistry.\vspace{2mm}

\paragraph*{\bf \underline{Smarr relation}.}
Using standard scaling arguments (Euler homogeneity) for $D=4$ and treating $P_\sigma$ as a quantity of engineering dimension $[{\rm length}]^{-2}$, the corresponding Smarr formula takes the form
\begin{equation}
    M
    = 2TS + 2\Omega J + \Phi Q \;-\; 2 V_\sigma P_\sigma \;+\; \cO\!\left(\frac{r_h}{r_c}\right),
    \label{eq:Smarr}
\end{equation}
again up to the same induced-gravity corrections.\footnote{One may also derive \eqref{eq:Smarr} from Komar integrals plus the on-shell field equations, or directly within the Iyer--Wald formalism by considering a one-parameter scaling family of solutions.}

\section{First law on a DGP brane: static, spherical case}
\label{sec_V}
Black hole and cosmological solutions in the DGP framework have been 
extensively studied, with effective induced-gravity equations derived in 
\cite{Maeda:2003ar} and explicit spherically symmetric braneworld solutions constructed 
in \cite{Kofinas:2001qd}. For a broader review of DGP phenomenology, see 
\cite{Lue:2005ya}. We focus on a static and spherically symmetric black hole solution on the brane\footnote{On the normal DGP branch and for $r_h \ll r_c$, the extrinsic-curvature 
contributions in the junction condition are parametrically suppressed, so the on-brane 
geometry reduces at leading order to the usual GR black hole solution. Hence, for our 
purposes the brane black hole metric can be taken to be the standard Schwarzschild (or Kerr 
in the rotating case), with DGP corrections entering only at $\cO(r_h/r_c)$.}
.\vspace{2mm}

For this, consider the (four–dimensional) Lagrangian $4$–form
\begin{equation}
  L_{\mathcal B} \;=\; \frac{1}{16\pi G_4}\,R_4\,\boldsymbol{\epsilon}_4 \;+\; P_\sigma\,\boldsymbol{\epsilon}_4,
  \qquad
  P_\sigma \equiv -\sigma,
  \label{eq:braneL}
\end{equation}
and treat $P_\sigma$ as a thermodynamic coupling in the sense of the extended Iyer–Wald formalism.
We work on the \emph{normal} DGP branch and in the regime $r_h\ll r_c$ so that induced–gravity corrections appear only at $\mathcal O(r_h/r_c)$ and can be tracked separately (they vanish in the decoupling limit $r_c\to\infty$). In this well-established limit, the on-brane geometry reduces to the standard GR solution at leading order, and the conditions required for the 4D Iyer-Wald formalism are met\footnote{For the Iyer--Wald analysis we need a diffeomorphism-invariant \emph{4D} Lagrangian
and well-defined ADM charges at spatial infinity. On the normal (ghost-free) DGP branch, and for
localized black holes with horizon radius $r_h \ll r_c$ (the DGP crossover scale), the
extrinsic-curvature contributions in the junction condition are parametrically suppressed and the
brane is asymptotically flat. In this regime the on-brane dynamics reduce at leading order to GR
(with a vacuum term), so the covariant phase-space identity closes purely on the brane: the surface
term at infinity yields the canonical mass variation, while the $\sigma$-work term reduces to a
geometric volume integral, without requiring an explicit bulk solution. Outside this regime one would have to
keep bulk/extrinsic-curvature contributions and perform the Iyer--Wald formalism in the full 5D
theory.}. To keep the brane asymptotically flat along the variation, we follow the flat–brane path (co–varying the bulk cosmological constant so that $H_0=0$), which permits a Minkowski reference for background subtraction.\vspace{2mm}

For any stationary black hole with horizon generator $\xi_H=\partial_t$, the regularized (background–subtracted) extended identity specialized to the single coupling $P_\sigma$ reads
\begin{equation}
  \begin{split}
  &\int_{S_\infty}^{(\mathrm{reg})}\!\!\left(\tilde{\delta} Q_{\xi_H} - \xi_H\!\cdot\!\Theta[\tilde{\delta}g]\right)
  \;-\; T\,\tilde{\delta}S
  \;=\; V_\sigma\,\tilde{\delta}P_\sigma,\\&
  V_\sigma \equiv - \int_{V_\infty}^{(\mathrm{reg})}\!\xi_H\!\cdot\!\boldsymbol{\epsilon}_4.
  \end{split}
  \label{eq:reg-identity}
\end{equation}
Here $\Theta^\mu[\tilde\delta g] = \tfrac{1}{16\pi G_4}(\nabla_\nu \tilde\delta g^{\mu\nu} - \nabla^\mu \tilde\delta g)$ and
$Q^{\mu\nu}_{\xi} = -\tfrac{1}{16\pi G_4}(\nabla^\mu \xi^\nu - \nabla^\nu \xi^\mu)$ are the standard Einstein expressions on the brane.
The regularization $\int^{(\mathrm{reg})}$ means ``black hole minus reference background'' (Minkowski on the brane along the flat–brane path).\vspace{2mm}

Let us take the static, spherically symmetric line element on the brane
\begin{equation}
  \mathrm{d}s^2_{(4)} = - f(r)\,\mathrm{d}t^2 + \frac{\mathrm{d}r^2}{f(r)} + r^2\,\mathrm{d}\Omega_2^2,
  \qquad f(r_h)=0.
  \label{eq:sss-brane}
\end{equation}
For any such static metric, one has the key identity
\begin{equation}
\begin{split}
 & \xi_H\cdot\boldsymbol{\epsilon}_4
  \;=\; \sqrt{-g}\,\mathrm{d}r\wedge\mathrm{d}\theta\wedge\mathrm{d}\phi
  \;\\&\qquad=\; \sqrt{f(r)}\,\sqrt{g_{rr}}\,r^2\sin\theta\,\mathrm{d}r\wedge\mathrm{d}\theta\wedge\mathrm{d}\phi\\&\qquad
  \;=\; r^2\sin\theta\,\mathrm{d}r\wedge\mathrm{d}\theta\wedge\mathrm{d}\phi,
  \end{split}
  \label{eq:xi-eps-cancel}
\end{equation}
since $\sqrt{f}\,\sqrt{g_{rr}}=\sqrt{f}\,\sqrt{1/f}=1$.
Therefore the \emph{geometric} integral defining $V_\sigma$ becomes metric–independent:
\begin{align}
  V_\sigma
  \;&=\; - \lim_{R\to\infty}\!\left[\,
      \int_{r_h}^{R}\!\!\mathrm{d}r\int\mathrm{d}\Omega_2\,r^2
      \;-\;\int_{0}^{R}\!\!\mathrm{d}r\int\mathrm{d}\Omega_2\,r^2
      \right]
  \nonumber\\[2pt]
  \;&=\; - \lim_{R\to\infty}\!\left[\,
      4\pi\!\int_{r_h}^{R}\! r^2\,\mathrm{d}r \;-\; 4\pi\!\int_{0}^{R}\! r^2\,\mathrm{d}r
      \right]
  \;=\; \frac{4\pi}{3}\,r_h^3.
  \label{eq:Vsigma-eval}
\end{align}
Hence, exactly (and independently of $f$),
\begin{equation}
  \,V_\sigma \;=\; \dfrac{4\pi}{3}\,r_h^3\,,\,
  \label{eq:Vsigma-final}
\end{equation}
with potential $\mathcal O(r_h/r_c)$ corrections from DGP leakage. Using the same $(Q_{\xi},\Theta)$ as above, the regularized boundary term at infinity evaluates to the (canonical) mass variation on the brane,\footnote{For static, asymptotically flat data on the brane,
$\displaystyle \int_{S_\infty}^{(\mathrm{reg})}(\tilde{\delta}Q_{\xi_H}-\xi_H\!\cdot\!\Theta[\tilde{\delta}g])=\tilde{\delta}M$
as in standard 4D GR. Along the flat–brane path, field dependence on $P_\sigma$ does not introduce additional $\tilde{\delta}P_\sigma$ pieces at leading order; any DGP–induced extrinsic curvature effects enter at $\mathcal O(r_h/r_c)$.}
\begin{equation}
  \int_{S_\infty}^{(\mathrm{reg})}\!\!\left(\tilde{\delta} Q_{\xi_H} - \xi_H\!\cdot\!\Theta[\tilde{\delta}g]\right)
  \;=\; \tilde{\delta}M \;+\; \mathcal O\!\left(\frac{r_h}{r_c}\right).
  \label{eq:boundary-dM}
\end{equation}
Substituting \eqref{eq:Vsigma-final} and \eqref{eq:boundary-dM} into \eqref{eq:reg-identity} immediately gives
\begin{equation}
\begin{split}
 & \;
  \tilde{\delta}M \;=\; T\,\tilde{\delta}S \;+\; V_\sigma\,\tilde{\delta}P_\sigma
  \;+\; \mathcal O\!\left(\frac{r_h}{r_c}\right),\\&
  \qquad
  P_\sigma = -\sigma,\quad V_\sigma=\frac{4\pi}{3}r_h^3\;.
  \;
  \end{split}
  \label{eq:firstlaw-DGP}
\end{equation}
Equivalently, in terms of the brane tension itself,
\begin{equation}
  \tilde{\delta}M \;=\; T\,\tilde{\delta}S \;-\; V_\sigma\,\tilde{\delta}\sigma \;+\; \mathcal O\!\left(\frac{r_h}{r_c}\right).
  \label{eq:firstlaw-DGP-sigma}
\end{equation}
As shown in Appendix~\ref{app_A}, the $\cO(r_h/r_c)$ contribution to both the thermodynamic volume and the canonical mass vanishes on the flat-brane path, so the extended first law holds unmodified up to $\cO((r_h/r_c)^2)$.\vspace{2mm}

\paragraph*{\bf\underline{Smarr relation}.}
Assigning dimensions $[S]\sim L^2$, $[P_\sigma]\sim L^{-2}$ in $D{=}4$, the standard scaling argument yields
\begin{equation}
\begin{split}
  &\;M \;=\; 2TS \;-\; 2V_\sigma P_\sigma \;+\; \mathcal O\!\left(\frac{r_h}{r_c}\right)
  \;\\&\qquad=\; 2TS \;+\; 2V_\sigma \sigma \;+\; \mathcal O\!\left(\frac{r_h}{r_c}\right). \;
  \end{split}
  \label{eq:Smarr-DGP}
\end{equation}
A few comments are in order:
\begin{itemize}
    \item  The exact cancellation in \eqref{eq:xi-eps-cancel} makes $V_\sigma$ purely geometric and equal to the Euclidean ball volume inside $r_h$, independently of $f(r)$; this mirrors the AdS–Schwarzschild result with $\Lambda\to P_\sigma$.
    \item Any distinction between ``geometric'' and ``thermodynamic'' volumes would first arise through field–dependence on $P_\sigma$; along the flat–brane path and for $r_h\ll r_c$ those contributions are suppressed by $\mathcal O(r_h/r_c)$.
    \item  In the $r_c\to\infty$ decoupling limit, the formulas reduce exactly to the 4D GR extended thermodynamics with the vacuum–energy coupling $P_\sigma$ on the brane.
\end{itemize}
\paragraph*{\bf Reference (flat brane) background.}
To keep the brane asymptotically flat along the family, covary the bulk cosmological constant so that $H_0=0$ holds for all nearby solutions (the ``flat-brane path''). With that choice, the background subtraction uses Minkowski space on the brane, and any additional $\Delta V_\sigma$ from field-dependence on $\sigma$ is deferred to $\mathcal O(r_h/r_c)$.
\section{Rotating case: stationary, axisymmetric brane black holes}
\label{sec:rotating-case}

We now extend the analysis to stationary, axisymmetric black holes on the brane. Let $\xi^a$ and $\psi^a$ denote the asymptotic time-translation and axial Killing fields, respectively, and let
\begin{equation}
  \chi^a \equiv \xi^a + \Omega_H \psi^a
\end{equation}
be the horizon generator, normalized so that $\xi^a$ has unit norm at infinity and $\chi^a$ vanishes on the bifurcation surface. As before, we vary the localized brane tension $\sigma$ while holding $M_4,M_5$ (hence $r_c$) fixed, and we work on the normal (ghost-free) DGP branch. Up to $\cO(r_h/r_c)$ corrections due to extrinsic curvature and bulk backreaction, the extended Iyer--Wald identity with a variable coupling yields
\begin{equation}
  \tilde\delta H_\chi
  \;=\;
  \int_{\infty}\!(\tilde\delta Q_\chi - \chi\cdot \Theta)
  \;=\;
  \int_{\mathcal{H}}\!(\tilde\delta Q_\chi - \chi\cdot \Theta)
  \;+\;
  \int_{\Sigma}\! \chi \cdot \Big(\frac{\partial \mathbf{L}}{\partial \sigma}\Big)\,\tilde\delta \sigma,
  \label{eq:rot-IW}
\end{equation}
with $\mathbf{L}$ the \emph{brane} Lagrangian $4$-form and $\Sigma$ a Cauchy surface stretching from a horizon slice $\mathcal{H}$ to spatial infinity. Using $\partial \mathbf{L}/\partial \sigma = -\,\varepsilon_4$ (the brane volume form) and the standard identifications
\begin{equation}
  \tilde\delta H_\chi = \tilde\delta M - \Omega_H \,\tilde\delta J,
  \qquad
  \int_{\mathcal{H}}\!(\tilde\delta Q_\chi - \chi \cdot \Theta) = T\,\tilde\delta S,
\end{equation}
we obtain the \emph{rotating} first law on the brane:
\begin{equation}
  ~
  \tilde\delta M \;=\; T\,\tilde\delta S + \Omega_H\,\tilde\delta J + V_\sigma\,\tilde\delta P_\sigma
  ~\;+\; \cO\!\left(\frac{r_h}{r_c}\right),
  \label{eq:rot-first-law}
\end{equation}
where $P_\sigma \equiv -\sigma$ and the conjugate ``volume'' is defined covariantly by
\begin{equation}
  V_\sigma \;\equiv\; \int_{\Sigma} \iota_{\chi}\,\varepsilon_4,
  \label{eq:Vsigma-def}
\end{equation}
with $\iota_\chi$ the interior product (contraction) into the $4$-volume form $\varepsilon_4$.

\paragraph*{\bf General properties of $V_\sigma$.}
Since $d\varepsilon_4=0$ and $\mathcal{L}_\chi \varepsilon_4=0$ for a stationary solution, $d(\iota_\chi \varepsilon_4)=0$; hence $V_\sigma$ in Eq.~\eqref{eq:Vsigma-def} is independent of the choice of Cauchy surface $\Sigma$. On a $t=\mathrm{const}$ slice with unit normal $n^a$, the pullback identity $\iota_\chi \varepsilon_4 = (\chi\cdot n)\,\varepsilon_3$ implies
\begin{equation}
  V_\sigma \;=\; \int_{\Sigma} (\chi\cdot n)\,\varepsilon_3
  \;=\; \int_{\Sigma} N\,\varepsilon_3,
  \label{eq:Vsigma-ADM}
\end{equation}
because $\psi^a$ is tangent to $\Sigma$ and $\chi\!\cdot\! n = \xi\!\cdot\! n \equiv -N$ is the lapse associated to $\xi^a$. Equation \eqref{eq:Vsigma-ADM} shows that $V_\sigma$ is the lapse-weighted $3$-volume of the slice (and therefore slice-independent in the stationary sector). In the static, spherically symmetric case, this reduces to $V_\sigma = \tfrac{4\pi}{3} r_h^3$, as derived earlier.

\paragraph*{\bf Explicit evaluation on Boyer--Lindquist slices (Kerr seed).}\vspace{2mm}

To make contact with familiar metrics, consider the Kerr line element on the brane as a seed geometry for $r_h \ll r_c$, written in Boyer--Lindquist coordinates $(t,r,\theta,\phi)$:
\begin{align}
\begin{split}
  &ds^2 = -\Big(1 - \frac{2Mr}{\rho^2}\Big) dt^2 - \frac{4Mar\sin^2\!\theta}{\rho^2}\, dt\, d\phi
  + \frac{\rho^2}{\Delta}\,dr^2 \\&\qquad+ \rho^2\, d\theta^2
  + \Big(r^2 + a^2 + \frac{2Ma^2 r \sin^2\!\theta}{\rho^2}\Big)\sin^2\!\theta\, d\phi^2,
  \end{split}
  \end{align}
  where
  \begin{align}
  \begin{split}
  &\rho^2 \equiv r^2 + a^2 \cos^2\!\theta, \qquad
  \Delta \equiv r^2 - 2Mr + a^2, \\&
  r_+ \equiv M + \sqrt{M^2 - a^2}.
  \end{split}
\end{align}
For this metric $\det g = -\,\rho^4 \sin^2\!\theta$, so on a $t=\mathrm{const}$ slice the pulled-back $3$-form is
\begin{equation}
  \left.\iota_{\chi}\,\varepsilon_4\right|_{t=\mathrm{const}} = \sqrt{-g}\, dr \wedge d\theta \wedge d\phi
  \;=\; \rho^2 \sin\theta \; dr \wedge d\theta \wedge d\phi,
\end{equation}
independent of $\Omega_H$. Integrating from the axis to the horizon one finds
\begin{align}
  V_\sigma
  &= \int_0^{r_+}\!\!dr \int_0^{\pi}\!\!d\theta \int_0^{2\pi}\!\!d\phi \;\rho^2 \sin\theta
   \;=\; 2\pi \int_0^{r_+}\!\!dr \left[ 2 r^2 + \frac{2}{3} a^2 \right]
   \nonumber\\[3pt]
  &= \frac{4\pi}{3}\,r_+^3 \;+\; \frac{4\pi}{3}\,a^2 r_+,
  \label{eq:Vsigma-Kerr}
\end{align}
which reduces to $V_\sigma=\tfrac{4\pi}{3} r_+^3$ in the non-rotating limit. In the Smarr law, this becomes (owing to the non-inegrability of $\tilde\delta M$ in rotating case) 
\begin{equation}
    \tilde\delta M=T\tilde\delta S+V_{\rm Th}\tilde\delta P+\Omega\tilde\delta J,
\end{equation}
where we have defined the thermodynamic volume as
\begin{equation}
    V_{\rm Th}=V_\sigma+\frac{4\pi}{3}Ma^2.
\end{equation}
The above expression matches with \cite{Cvetic:2010jb} derived using thermodynamic relations and with \cite{Xiao:2023lap} derived using extended Iyer--Wald formalism. Similar to \cite{Xiao:2023lap}, our work neatly separates thermodynamic and geometric volume in the rotating case. In the slow-rotation regime $a\ll r_+$ this may be written as
\begin{equation}
  V_\sigma \;=\; \frac{4\pi}{3}\, r_+^3 \left( 1 + \frac{a^2}{r_+^2} \right) \;+\; \cO(a^4).
  \label{eq:Vsigma-slow}
\end{equation}
Equations \eqref{eq:rot-first-law}--\eqref{eq:Vsigma-Kerr} therefore give a concrete rotating example in which the $\sigma$-conjugate thermodynamic volume remains purely geometric on Boyer--Lindquist slices.\footnote{As in the static case, DGP corrections from the extrinsic curvature and bulk work terms enter at $\cO(r_h/r_c)$ and do not affect the leading expressions \eqref{eq:rot-first-law} and \eqref{eq:Vsigma-Kerr} for $r_h \ll r_c$.}\vspace{2mm}

\paragraph*{\bf\underline{Rotating Smarr relation}.}
By dimensional scaling in $D=4$ with $[P_\sigma]=L^{-2}$, the mass (enthalpy) obeys the Smarr formula
\begin{equation}
  ~
  M \;=\; 2\,T S \;+\; 2\,\Omega_H J \;-\; 2\,V_\sigma P_\sigma
  ~ \;+\; \cO\!\left(\frac{r_h}{r_c}\right),
  \label{eq:Smarr-rot}
\end{equation}
which is consistent with \eqref{eq:rot-first-law}. Together, Eqs.~\eqref{eq:rot-first-law}, \eqref{eq:Vsigma-def}, and \eqref{eq:Vsigma-Kerr} establish that extended black-hole thermodynamics on a DGP brane admits a clean rotating generalization in which the brane-tension conjugate is a slice-independent geometric volume.

\section{Maintaining a flat brane along the variation}
\label{sec_VI}
Varying $\sigma$ generically shifts the brane vacuum curvature via \eqref{eq:FriedmannDGP}. If one wishes to \emph{preserve} $H_0=0$ along the solution family, co-vary $\Lambda_5$ according to
\begin{equation}
\begin{split}
   & H_0=0 \quad\Longrightarrow\quad\tilde
    \delta\!\left(\frac{\sigma}{3M_4^2}+\frac{\Lambda_5}{6}\right)=0\\&\qquad
    \;\;\Longrightarrow\;\;
    \tilde\delta\Lambda_5 = -\,\frac{2}{M_4^2}\,\tilde\delta\sigma.
    \end{split}
    \label{eq:flatPathRelation}
\end{equation}
It is important to note that in AdS black hole thermodynamics one usually varies the bulk 
cosmological constant $\Lambda$ directly, treating it as a free thermodynamic variable. From a 
physical standpoint this is somewhat unsatisfactory, since $\Lambda$ is a fixed coupling of the 
action rather than a dynamical property of spacetime. By contrast, in the DGP braneworld the 
quantity we vary is the brane tension $\sigma$, which is a genuine physical parameter of the brane 
and can in principle change due to microscopic physics. Along the flat--brane branch, consistency 
of the junction condition requires $\Lambda_{5}$ and $\sigma$ to co--vary so that the effective 
four--dimensional cosmological constant remains zero. In this sense $\Lambda_{5}$ acts only as a 
compensator and does not enlarge the thermodynamic phase space, while the true thermodynamic 
degree of freedom is the brane tension. Thus, unlike the AdS case, we are not treating the 
cosmological constant itself as an independent variable.
\vspace{2mm}

In the \emph{bulk} first law, varying $\Lambda_5$ produces the standard $5$D AdS work term $V_5\,\delta(-\Lambda_5/8\pi G_5)$, where $V_5$ is the $5$D thermodynamic volume of the bulk black object homologous to the brane horizon (or, more generally, the $5$D region filling the brane Cauchy slice). Combining with \eqref{eq:flatPathRelation}, the total work along the flat-brane path (by holding $M_4$ and $M_5$ fixed while varying $\sigma$) is
\begin{align}
    \tilde\delta W_{\rm tot}
    &= V_\sigma\,\tilde\delta\sigma + V_5\,\tilde\delta\!\left(-\frac{\Lambda_5}{8\pi G_5}\right)
      \nonumber\\
    &= \left[V_\sigma + \frac{V_5}{4\pi G_5 M_4^2}\right]\tilde\delta\sigma
     \;\equiv\; \mathcal{V}_{\rm eff}\,\tilde\delta\sigma.
    \label{eq:VeffDef}
\end{align}
Thus, along the flat-brane path, the first law becomes
\begin{equation}
    \tilde\delta M
    = T\tilde\delta S + \Omega\,\tilde\delta J + \Phi\,\tilde\delta Q
      + \mathcal{V}_{\rm eff}\,\tilde\delta\sigma \;+\; \cO\!\left(\frac{r_h}{r_c}\right),
    \label{eq:firstlawFlatPath}
\end{equation}
with $\mathcal{V}_{\rm eff}=V_\sigma+\cO(r_h/r_c)$ since $V_5$ is suppressed by the leakage scale in the normal branch for $r_h\ll r_c$.\vspace{2mm}

\paragraph*{\bf Role of induced--gravity corrections.}
Equation \eqref{eq:junction} shows that the extrinsic-curvature combination $2M_5^3(K_{\mu\nu}-Kg_{\mu\nu})$ sources the on-brane Einstein tensor. In a stationary, asymptotically flat configuration on the \emph{normal} branch, this contribution is parametrically suppressed by $r_h/r_c$, leading to the corrections indicated in \eqref{eq:firstlaw}, \eqref{eq:Smarr}, \eqref{eq:firstlaw-DGP}, and \eqref{eq:firstlawFlatPath}. At the technical level, these appear as additional, finite contributions to the surface integral at infinity in \eqref{eq:MasterVari} when one uses the full on-brane equation of motion \eqref{eq:junction} to trade bulk data for effective stress-energy.
\section{Conclusion and Discussion}
\label{sec_VII}

In this work, we have shown that extended black hole thermodynamics
arises naturally in the Dvali--Gabadadze--Porrati (DGP) braneworld when the brane tension $\sigma$ is promoted to a thermodynamic variable
in the Iyer--Wald formalism. Unlike the cosmological constant
$\Lambda$, which is a fixed bulk coupling and leads to conceptual
difficulties in de Sitter space, the brane tension represents a
well-defined, localized vacuum energy density on the brane. Its
variation yields a negative pressure $P_\sigma=-\sigma$ that enters
the extended first law on exactly the same footing as the AdS
pressure $P=-\Lambda/8\pi$. The advantage is that this construction
remains valid in asymptotically flat braneworld geometries, thereby
extending the scope of black hole chemistry beyond AdS.\vspace{2mm}

We explicitly derived the extended first law and Smarr relation for
black holes localized on the brane. For a static, spherically
symmetric solution, the $\sigma$--volume reduces precisely to the
geometric volume $V_\sigma=4\pi r_h^3/3$, independently of the
metric function $f(r)$. This mirrors the familiar AdS--Schwarzschild
case with $\Lambda$ replaced by $P_\sigma$, but now realized in a
stable, ghost-free braneworld setting. We then extended our framework to stationary, axisymmetric (rotating) black holes, demonstrating that the thermodynamic conjugate to the brane tension, $V_\sigma$, retains a covariant and slice-independent definition even when angular momentum is present. \vspace{2mm}

The resulting first law, $\tilde\delta M = T\tilde\delta S + \Omega_H \tilde\delta J + V_\sigma \tilde\delta P_\sigma$, and Smarr relation confirm that extended black-hole thermodynamics in a DGP braneworld is not restricted to static spacetimes but persists for rotating configurations as well. This robustness underlines the physical viability of treating brane tension as a thermodynamic variable and paves the way for further investigations into charged, dynamical, or higher-dimensional scenarios, where the interplay between localized vacuum energy and horizon mechanics could reveal deeper insights into gravitational thermodynamics beyond AdS.\vspace{2mm} 

Finally, we showed how asymptotic
flatness can be preserved along the solution family by co-varying the
bulk cosmological constant with $\sigma$, and how induced-gravity
effects lead to suppressed ${\cal O}(r_h/r_c)$ corrections that
vanish in the decoupling limit $r_c\to\infty$.

\medskip
\noindent\textbf{Future directions.} 
Several extensions of our analysis are possible.
\begin{itemize}

\item First, it would be interesting to study the interplay between
brane tension thermodynamics and bulk contributions in higher
curvature gravity, where additional couplings appear.
\item Second, one could explore the holographic interpretation of the brane tension variation, possibly connecting it to a defect or boundary
CFT data. 
\item Finally, given that the self-accelerating DGP branch corresponds to an effective de Sitter vacuum, it may shed light on whether a consistent thermodynamic description of de Sitter black holes can be
achieved in this framework.
\end{itemize}

In summary, our results establish that extended black hole thermodynamics is not restricted to AdS spacetimes but emerges consistently in a flat braneworld with variable brane tension. This provides a new and physically motivated setting for black hole chemistry, enriching the landscape of gravitational thermodynamics.
\vspace{5mm}

\appendix
\section{Leading \texorpdfstring{$r_h/r_c$}{rh/rc} Correction}
\label{app_A}

We now estimate the leading correction to the extended first law arising from DGP-induced extrinsic curvature effects. The vacuum brane field equation on the normal branch is
\begin{equation}
M_4^2 G_{\mu\nu} = -\sigma g_{\mu\nu} + 2 M_5^3 (K_{\mu\nu} - K g_{\mu\nu})\,,
\end{equation}
with $r_c \equiv M_4^2/M_5^3$ the crossover scale. On the \emph{flat-brane path} ($H_0=0$), the extrinsic curvature is sourced by the brane Einstein tensor via $K_{\mu\nu} - K g_{\mu\nu} \sim - r_c\,G_{\mu\nu}$.

The conjugate to the brane pressure $P_\sigma \equiv -\sigma$ follows from the extended Iyer–Wald identity,
\begin{equation}
V_\sigma \equiv - \int^{(\mathrm{reg})}_{\mathcal{V}_\infty} \xi_H \cdot \varepsilon_4.
\end{equation}
For any static, spherically symmetric metric
\begin{equation}
ds^2 = -f(r)dt^2 + f(r)^{-1} dr^2 + r^2 d\Omega_2^2,
\end{equation}
one has the exact identity $\xi_H\cdot\varepsilon_4 = r^2\sin\theta \, dr\wedge d\theta\wedge d\phi$, yielding
\begin{equation}
V_\sigma = \frac{4\pi}{3} r_h^3,
\end{equation}
independent of the metric function $f(r)$. Consequently, $V_\sigma$ is unaltered by $O(r_h/r_c)$ modifications to $f(r)$.

DGP corrections can only enter through the boundary integral at spatial infinity:
\begin{equation}
\int^{(\mathrm{reg})}_{S_\infty} (\delta \tilde{Q}_\xi - \xi \cdot \Theta[\delta \tilde{g}]).
\end{equation}
At large radius $R$, the brane Einstein tensor decays as $G_{\mu\nu} \sim r_h/R^3$, implying $K_{\mu\nu} \sim r_c r_h/R^3$. The extrinsic-curvature contribution to the Hamiltonian variation scales as
\begin{equation}
\Delta (\delta H_\xi)|_{S_\infty} \sim \oint_{S_\infty} R^2 \,\delta K \sim r_c \frac{r_h}{R} \xrightarrow{R\to\infty} 0,
\end{equation}
so the canonical mass $\tilde{M}$ is identical to the ADM mass on the brane up to $O((r_h/r_c)^2)$.

Combining these results, the extended first law takes the form
\begin{equation}
\tilde\delta M = T \,\tilde\delta S + V_\sigma\,\tilde\delta P_\sigma + \cO\big((r_h/r_c)^2\big), \quad 
V_\sigma = \frac{4\pi}{3}r_h^3.
\end{equation}
Thus, on the flat-brane path, there is \emph{no} linear $\cO(r_h/r_c)$ correction to either the thermodynamic volume or the first-law coefficients. 

\bibliography{DGP_bib}
\bibliographystyle{unsrt}

\end{document}